\documentclass[np2,twoside,onecolumn,showpacs]{revtex4}

\usepackage{amssymb}
\usepackage{amsmath}
\usepackage{bm,multirow}
\usepackage[pdftex]{graphicx}
\usepackage[pdftex,colorlinks = true,
                      linkcolor = blue,
                      urlcolor = blue,
                      citecolor = blue,
                      anchorcolor = blue]{hyperref}

\begin{document}

\title{Video recognition by physical reservoir computing in magnetic materials}

\author{Kaito \surname{Kobayashi}}
\email{kaito-kobayashi92@g.ecc.u-tokyo.ac.jp}
\author{Yukitoshi \surname{Motome}}
\affiliation{Department of Applied Physics, University of Tokyo,
Bunkyo-ku, Tokyo, 113-8656, Japan}

\begin{abstract}
	Nonlinear spin dynamics in magnetic materials offers a promising avenue for implementing physical reservoir computing, one of the most accomplished brain-inspired frameworks for information processing. 
	In this study, we investigate the practical utility of magnetic physical reservoirs by assessing their performance in a video recognition task. 
	Leveraging a recently developed spatiotemporal parallelization scheme, our reservoir achieves accurate classifications of previously provided images. 
	Our findings pave the way for the development of visual sensors based on the magnetic physical reservoir computing. 
\end{abstract}


\maketitle

\section{Introduction}

Physical reservoir computing represents a state-of-art framework for information processing. 
This methodology harnesses the complex dynamics in a physical reservoir as an embodiment of the recurrent neural network, which nonlinearly projects input data onto a high-dimensional internal feature space. 
Optimization is limited to the read-out layer where the computational results are extracted. 
This reduction in training cost enables low-power and high-speed hardware implementations of real-time machine learning\cite{r1,r2}. 
Among the plethora of potential platforms, magnetic physical reservoirs have attracted considerable research interest due to their pronounced nonlinearity and high dimensionality arising from the complicated spin dynamics\cite{r3,r4}. 
Indeed, recent demonstrations of recognition tasks serve as strong evidence of their potential for practical applications beyond simple benchmarking tasks\cite{r4,r5,r6}. 
Nevertheless, to ensure an adequate quantity of nodes for the classification of large dimensional features, the majority of physical reservoirs allocate time degrees of freedom for processing distinct features\cite{r5,r6}. 
This approach does not effectively exploit the essence of physical reservoirs, namely the fading memory property\cite{r1}, because temporal degrees of freedom is not used for the retention of the short-term memory.
Hence, the capability of magnetic physical reservoirs to address practical recognition tasks involving temporal processing remains largely unexplored. 

In this paper, utilizing a prototypical frustrated magnet as our physical reservoir, we demonstrate a video recognition task that extends single-instance recognition over time. 
Here, by leveraging the recently developed spatiotemporal massive parallelization scheme\cite{r3}, we  simultaneously process each element of a large-dimensional image matrix without the need to partition time degrees of freedom. 
Consequently, we can take full advantage of the short-term memory, and our reservoir accurately recognizes images even in the presence of temporal delays. 
Our findings clearly affirm the competency of magnetic physical reservoirs even in practical temporal tasks, thus expanding a range of real-world applications. 

\section{Model and method}
We utilize an antiferromagnetic Heisenberg model on a \(L\times L\) \((L=128)\) triangular lattice with open boundary conditions as our physical reservoir. 
The Hamiltonian is given by 
\begin{equation}
	\mathcal{H}=J\sum_{\langle i,j \rangle}\mbox{\boldmath$S$}_i\cdot \mbox{\boldmath$S$}_j -\sum_{j\in \Lambda} H^z_{j}(t)S_{j}^z,
\end{equation}
where \(\mbox{\boldmath$S$}_i=(S_i^x,S_i^y,S_i^z)\) is the classical spin at site \(i\) and \(\langle i,j\rangle\) represents a pair of nearest-neighbor sites; \(H^z_{j}(t)\) is the magnetic field for the input and \(\Lambda\) is a set of lattice sites used as the input terminal spins. 
We take \(J\equiv 1\) as the energy unit, and the ground state without magnetic fields has an in-plane \(120^\circ\) structure.  

\begin{figure}[bhtp]
	\includegraphics[width=\hsize]{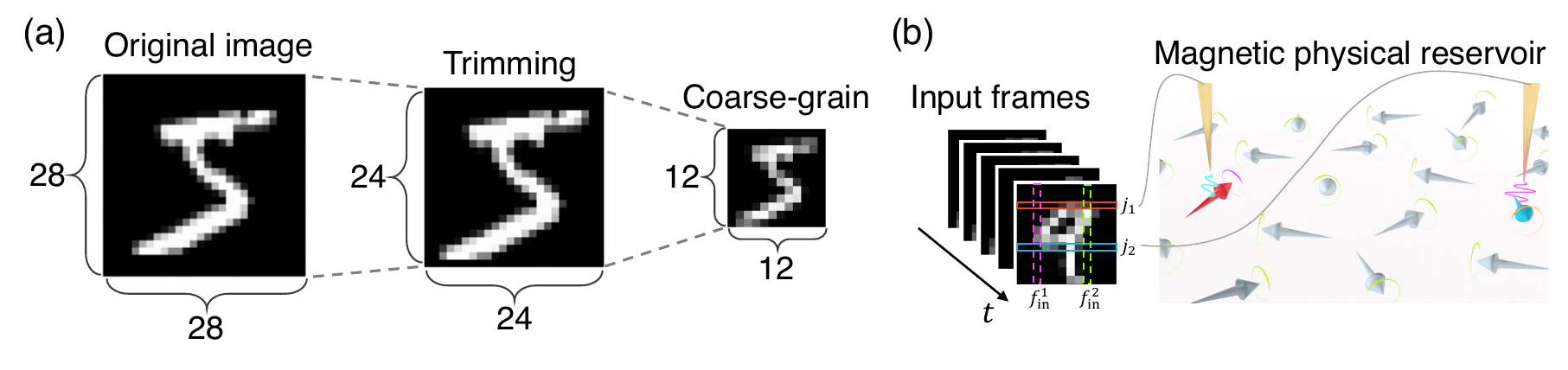}
	\caption{(a) Schematic of pre-processing to reduce a image size. 
	(b)	Concept of the video recognition task. 
	The images are successively provided into the magnetic physical reservoir using spatiotemporal parallelization.}
	\label{fig1}
\end{figure}

Here, we examine a video recognition task, which is an illustrative example of a practical task necessitating temporal memories. 
We sequentially provide handwritten digit images extracted from the MNIST database\cite{r7}, and the objective of this task is to accurately identify the digit written in the image \(d\) frames preceding the current one. 
A fine performance in the recognition of the image with the delay \(d>0\) indicates the capability to handle continuously evolving images, i.e. a video. 
When inputting the original \(28 \times 28\) pixel images, we trim the irrelevant area near the edge to \(24 \times 24\) pixels, and then coarse-grain the images into \(12 \times 12\) pixels by averaging each \(2 \times 2\) pixel area [Fig.~\ref{fig1}(a)]. 
Each image is thus considered as a \(12 \times 12\)-matrix \(M\), whose \((m,n)\) component, \(M_{mn}\), is in the range of \(0\leq M_{mn}\leq 1\). 
The digit depicted in the image is represented by means of a one-hot vector \(\bm{L}\). 
This is a \(10\)-dimensional vector, with all components having values of zero except the one corresponding to the digit \(l\) in the image, where \(L^{l+1}=1\). 

Following the parallelization protocol in both frequency domain and real space\cite{r3}, we prepare \(12\) distinct input frequencies, \(f^m_{\mathrm{in}}\), and \(12\) input terminal spins, \(j_n\), where \(m\) and \(n\) are both positive integers satisfying \(1\leq m, n \leq 12\). 
We take \(f_{\mathrm{in}}^m=k_m/(2t_{\mathrm{in}})\)  with an integer \(k_m\), which ensures the continuity of \(H_j^z (t)\) in time; each \(k_m\) is assigned to avoid being an integer multiple of any other. 
Similarly, in order to reduce crosstalk, \(j_n\) are carefully positioned more than \(24\) spins apart. 
The image \(M\) is then converted into the input magnetic fields as 
\begin{equation}
	\label{MagneticField}
	H^z_{j_n}(t)=\sum_{m=1}^{12} H_{\mathrm{in}} \sin\left(2\pi f^m_{\mathrm{in}}\right)\times(M_{mn}-0.5),
\end{equation}
where \(H_{\mathrm{in}}\) is the norm of magnetic field [Fig.~\ref{fig1}(b)]. 

At zero temperature, the spin dynamics under the input field 
is simulated by the Landau-Lifshitz-Gilbert (LLG) equation
\begin{equation}
  \frac{d\mbox{\boldmath$S$}_i}{dt}=-\frac{1}{1+\alpha^2}\left[\mbox{\boldmath$S$}_i\times\mbox{\boldmath$H$}_i^\mathrm{eff}+\alpha\mbox{\boldmath$S$}_i\times(\mbox{\boldmath$S$}_i\times\mbox{\boldmath$H$}_i^\mathrm{eff})\right],
\end{equation}
where \(\alpha\) is the Gilbert damping constant and 
\(\mbox{\boldmath$H$}_i^\mathrm{eff} = -\partial \mathcal{H}/\partial \bm{S}_i\) is the effective magnetic field at site \(i\). 
We prepare a total of \(17,000\) images and successively provide to the reservoir with an interval of time, \(t_{\mathrm{in}}\), by changing the input magnetic field based on the input image matrix as Eq.~(\ref{MagneticField}). 
In the following calculations, 
we take \(H_{\mathrm{in}}=0.5\), \(\alpha = 0.1\) and \(t_{\mathrm{in}}=12\). 

Under the input of the \(k\)-th image \(M_k\) with the label vector \(\bm{L}_k\), the internal state of the reservoir is represented by an \((N_t+1)\)-dimensional vector, denoted as \(\bm{X}_k\), which is given by
\begin{equation}
	\mbox{\boldmath$X$}_k = \left(S^z_{\lambda}(kt_{\mathrm{in}}),S^z_{\lambda}(kt_{\mathrm{in}}+\Delta t_{\mathrm{in}}),S^z_{\lambda}(kt_{\mathrm{in}}+2\Delta t_{\mathrm{in}}),\dots , S^z_{\lambda}(kt_{\mathrm{in}}+(N_{\mathrm{t}}-1)\Delta t_{\mathrm{in}}),1\right)
\end{equation}  
where \(\Delta t_{\mathrm{in}}=t_{\mathrm{in}}/N_t\) and we take \(N_t = 120\). 
By linearly transforming the internal state vector with a \([(N_t+1)\times10]\)-dimensional weight matrix \(W\), the output label vector \(\bm{y}_k\) is calculated as \(\bm{y}_k = \mbox{\boldmath$X$}_k W\). 
The weight matrix \(W\) is trained based on the least squared method so that each \(\bm{y}_k\) becomes close to the desired output \(\bar{\bm{y}}_k\); in the video recognition task, we take \(\bar{\bm{y}}_k = \bm{L}_{k-d}\) with the delay step \(d\). 
The output digit is determined as the index \(l'\) where \(y_k^{l'+1}\) is closest to \(1\). 
We used \(2,000\) images to warm up the system, \(12,000\)  for training, and \(3,000\)  for testing. 

\section{Results} 
\begin{figure}[htbp]
	\includegraphics[width=\hsize]{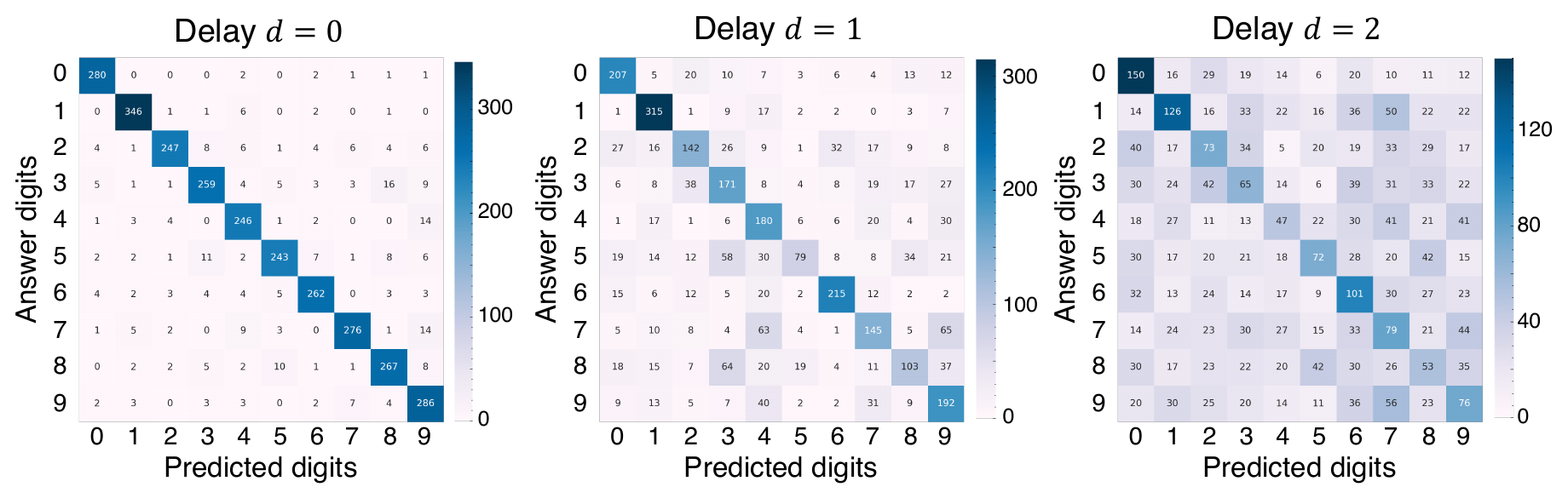}
	\caption{Confusion matrix describing the performance for the video recognition task with delay \(d=0,1\) and \(2\). 
	Each element shows the number of instances corresponding the prediction digits and the answer digits.}
	\label{fig2}
\end{figure}
Figure \ref{fig2} represents the confusion matrix for the video recognition task with delay \(d=0,1\) and \(2\). 
By definition, the video recognition with \(d=0\) is equivalent to the single-instance recognition. 
Remarkably, the diagonal components of the confusion matrix are dominant, whereas the off-diagonal components manifest negligible values. 
Indeed, the recognition accuracy, which is the ratio where the predicted digit is equal to the answer digit, is evaluated as \(90.4\%\). 
This observation supports the capability of our reservoir for precisely recognizing digits in the image matrix. 
It is worth noting that, despite the considerable simplicity of our reservoir system compared to magnetic skyrmions, the recognition accuracy is comparable to the previous results achieved by the skyrmion-based reservoirs, both in the simulation (\(87.6 \%\))\cite{r5} and in the experiment (\(94.7\pm 0.3\%\))\cite{r6}. 
There has been a conjecture that complex spin textures could enhance the nonlinearity and high dimensionality of magnetic physical reservoirs, thereby potentially improving the classification performance\cite{r3,r6}. 
Nonetheless, the comparable performance between  simple and complex magnets suggests the necessity for more comprehensive benchmarking tasks, beyond the single-instance recognition, to thoroughly evaluate the influence of detailed magnetic structures on information processing.

The core significance of the video recognition task lies in cases with \(d>0\), where our reservoir is required to classify images relying on temporal memories. 
In the case of \(d=1\), the diagonal components of the confusion matrix continue to dominate, although the increased prominence of errors. 
For the more extended delay of \(d=2\), many off-diagonal components become nonzero, yet the dominance of the diagonal components persists. 
The recognition accuracy is quantified as \(58.3\%\) for \(d=1\) and \(28.0\%\) for \(d=2\). 
Given that the accuracy is expected to be \(10\%\) in the case of random estimation, the improvement in accuracy indicates the ability of our reservoir to process images even in the presence of temporal delays. 
We emphasize that this video recognition critically relies on advancements in the massive parallelization scheme\cite{r3}. 
Otherwise, the allocation of time degrees of freedom would be imperative for loading a large dimensional image, limiting the system to single-instance recognitions without temporal delays.  

\section{Concluding remarks}

To summarize, we have explored the capabilities of magnetic physical reservoirs to address practical tasks involving temporal processing through the evaluation of the video recognition task. 
We have employed a spatiotemporal parallelization scheme to handle large-dimensional image matrices, avoiding the partitioning of time degrees of freedom while effectively harnessing the short-term memory. 
Consequently, our reservoir has exhibited remarkable accuracy in recognizing digits in the images, even when temporal delays are introduced. 
Furthermore,  we have observed comparable levels of performance in the single-instance recognition task regardless of the complexity of the magnetic texture. 
In order to investigate the impacts of magnetic interactions, it would be essential to carefully formulate dedicated tasks such that their performances reflect details of the magnetic properties. 
We leave the comprehensive elucidation of the mechanism underlying information processing capabilities in magnetic physical reservoirs to  future investigations. 

Our study highlights the immense potential of magnetic physical reservoirs to address practical tasks involving temporal delays. 
Particularly noteworthy is the prospect of leveraging their capability to process video data, facilitating the realization of visual sensors for real-time object detection and anomaly mitigation. 
In the ever-evolving landscape of information processing, our findings chart a clear path toward the creation of powerful tools through the application of magnetic physical reservoir computing.

\section*{ACKNOWLEDGEMENTS}
This work was supported by a Grant-in-Aid for Scientific Research on Innovative Areas “Quantum Liquid Crystals” (KAKENHI Grant No. JP19H05825) from JSPS of Japan and JST CREST (No. JP-MJCR18T2). 
K.K. was supported by the Program for Leading Graduate Schools (MERIT-WINGS). 
The computation in this work has been done using the facilities of the Supercomputer Center, the Institute for Solid State Physics, the University of Tokyo.


\begin{references}

\bibitem{r1} H. Jaeger and H. Haas, \href{http://dx.doi.org/10.1126/science.1091277}{Science \textbf{304}, 78-80 (2004).}

\bibitem{r2} G.	Tanaka, T. Yamane, J. B. Héroux, R. Nakane, N. Kanazawa \textit{et al.}, \href{http://dx.doi.org/10.1016/j.neunet.2019.03.005}{Neural Netw. \textbf{115}, 100-123 (2019).}

\bibitem{r3} K.	Kobayashi and Y. Motome, \href{https://doi.org/10.1038/s41598-023-41757-3}{Sci. Rep. \textbf{13}, 15123 (2023).}

\bibitem{r4} J. Torrejon, M. Riou, F. A. Araujo, S. Tsunegi, G. Khalsa \textit{et al.}, \href{http://dx.doi.org/10.1038/nature23011}{Nature \textbf{547}, 428-431 (2017).}

\bibitem{r5} W. Jiang, L. Chen, K. Zhou, L. Li, Q. Fu, \textit{et al.}, \href{http://dx.doi.org/10.1063/1.5115183}{Appl. Phys. Lett. \textbf{115}, 192403 (2019).}

\bibitem{r6} T. Yokouchi, S. Sugimoto, B. Rana, S. Seki, N. Ogawa \textit{et al.}, \href{http://dx.doi.org/10.1126/sciadv.abq5652}{Sci. Adv. \textbf{8}, eabq5652 (2022).}

\bibitem{r7} Y. LeCun, L. Bottou, Y. Bengio and P. Haffner, \href{http://dx.doi.org/10.1109/5.726791}{Proc. IEEE \textbf{86}, 2278-2324 (1998).}

\end{references}
\end{document}